# Weighted Soft Decision for Cooperative Sensing in Cognitive Radio Networks

Mohammad Iqbal Bin Shahid and Joarder Kamruzzaman
Gippsland School of IT, Monash University, VIC 3842, Australia
Email: <mohammad.iqbal.shahid, joarder.kamruzzaman>@infotech.monash.edu.au

*Abstract* - Enhancing the current services or deploying new services operating in RF spectrum requires more licensed spectrum which may not be provided by the regulatory bodies because of spectrum scarcity. On the contrary, recent studies suggest that many portions of the licensed spectrum remains unused or underused for significant period of time raising the issue of spectrum access without license in an opportunistic manner. Among all the spectrum accessing techniques, sensing based methods are considered optimal for their simplicity and cost effectiveness. In this paper, we introduce a new cooperative spectrum sensing technique which considers the spatial variation of secondary (unlicensed) users and each user's contribution is weighted by a factor that depends on received power and path loss. Compared to existing techniques, the proposed one increases the sensing ability and spectrum utilization, and offers greater robustness to noise uncertainty. Moreover, this cooperative technique uses very simple energy detector as its building block thereby reduces the cost and operational complexity.

## I. INTRODUCTION

The radio frequency spectrum is a limited resource with great importance. Deployment and expansion of services by diverse wireless service providers necessitates the increasing accommodation of services in this scarce resource. But the current policy of frequency allocation which is conducted by the government agencies by giving licenses to service providers is not efficient enough to meet this ever-mounting demand. On the other hand, recent investigations reveal the fact that licensed spectrum is greatly under-utilized [1]. This increasing demand for the spectrum contradicting the insufficiency of vacant bands necessitates a new spectrum policy. FCC has taken initiative to open up the TV bands for unlicensed access [2], IEEE has formed a working group (IEEE 802.22) [3] and many other major organizations are also working on this issue signifying an inevitable shift in the spectrum access policy [4]. The new spectrum policy incorporates the strategy where a secondary user group (unlicensed users) can access the temporarily unused licensed bands, i.e., white spaces of primary users (licensed users) on a non-interfering basis. Therefore, the best way to acquire the status of the primary user's licensed band is direct spectrum sensing [5]. In this process, secondary users can access the licensed spectrum only if there are no primary user activities on a particular band. Therefore, interference on the primary users is protected by utilizing only the white spaces by the secondary users instead of using the conventional interference protection methods.

Direct sensing based on energy detection has been proven to be a simple method for detecting the presence of primary transmitter. But when this method is applied in a stand-alone cognitive radio device, due to fading and/or shadowing it might draw a false conclusion about the presence of primary transmitter. For example, if a (secondary) user senses low or no primary user signal in a certain spectrum band, it may be due to shadowing or heavy fading rather than being a white space. To address these problems, cooperation between the users is extremely necessary.

In cooperative sensing, the uncertainty due to fading/ shadowing is minimized by accumulating the sensing results from different users and taking a combined decision about the presence of the primary licensee in the target licensed band. Based on this idea, cooperative spectrum sensing in cognitive radio networks has been studied intensely in recent literature [6]-[11]. Research on cooperative sensing is focused on several areas including the designing of optimal detector for sensing [7], optimal link budget [9], distance-user tradeoff in correlated fading environment [11] and devising new models and communication methods to detect primary transmitter [10]. But the most significant analyses are established on simple energy detection and combining methods [6], [8]. Energy detection has been proven to be a simple but effective method for detecting the presence of primary transmitter. But when this method is applied in a stand-alone cognitive radio device, due to fading and/or shadowing it might take incorrect decision about the presence of primary transmitter. In such cases, the cooperative spectrum sensing where the users take a combined decision based on their individual sensing results yields better detection than other sensing methods. However, improved co-operative scheme is needed to further enhance detection accuracy and spectrum utilization for practical deployment and wide-spread use of cognitive radio. In this regard this paper makes the following contributions: i) proposes a new combining method (weighted combining) that incorporates simple energy detector for user cooperation by considering the spatial variations of the users; ii) with the same number of cooperating users in independently faded channels, the proposed method detects the presence of primary user with higher probability than the existing combining methods [8]; iii) achieves higher spectrum utilization and elevated agility with lower observation time, bandwidth and SNR requirements and iv) requires less number of users to achieve a given detection probability.

The rest of this paper is organized as follows. Section II explains the energy detection method and local spectrum sensing and highlights the importance of cooperation. The cooperative sensing method is explained in Section III. In Section IV, the proposed weighted combining method is thoroughly described. The results are investigated, analyzed and compared with existing methods in Section V. Finally, this paper is concluded by providing some final remarks in Section VI.

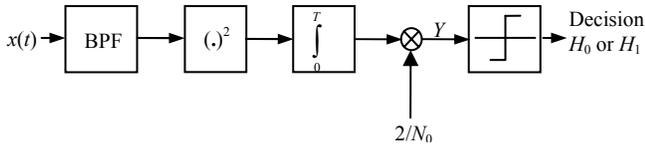

Fig. 1. Energy Detector

## II. ENERGY DETECTOR AND INDIVIDUAL SPECTRUM SENSING

Fig. 1 shows the block-diagram of an energy detector. The input bandpass filter with center frequency $f_c$ and bandwidth of interest $B$, filters the out-of-band noise. This filter is followed by a squaring device in cooperation with an integrator which measures the received energy over the observation interval, $T$. The output of the integrator is then normalized by $N_0/2$, where $N_0$ represents one-sided noise power spectral density. For individual sensing, the normalized output $Y$ is then compared to a decision threshold $\lambda$ to decide the presence of primary transmitter.

The objective of spectrum sensing is to determine whether a desired frequency band is currently being used by the primary licensee and can be formulated as a binary hypothesis testing problem-

$$x(t) = \begin{cases} v(t), & H_0 \ (White\ space) \\ g\,s(t) + v(t), & H_1 \ (occupied) \end{cases} \quad (1)$$

where $x(t)$ is the received signal by the (secondary) user, $s(t)$ is transmitted signal of primary transmitter, $v(t)$ is the additive white Gaussian noise (AWGN) and $g$ is the amplitude gain of the channel. The SNR can be defined as $\gamma = W/(N_0 B)$, where $W$ is the primary signal power received the user.

Individual sensing results for a user can be achieved analytically by the use of $Y$ (which acts as the decision statistic) and $\lambda$. The time-bandwidth product $TB$, denoted by $r$ is assumed to be an integer number for simplicity. $Y$ has central and non-central chi-square distributions under $H_0$ and $H_1$, respectively. These distributions has $2r$ degrees of freedom and the latter distribution has a noncentrality parameter of

$$\frac{WT}{N_0/2} = \frac{2WTB}{N_0 B} = 2TB\frac{W}{N_0 B} = 2r\gamma \quad (2)$$

The decision statistic can then be expressed as

$$Y \sim \begin{cases} \chi^2_{2r}, & H_0 \\ \chi^2_{2r}(2r\gamma), & H_1 \end{cases} \quad (3)$$

The probability density function (PDF) of $Y$ can then be expressed as [8]

$$f_Y(y) = \begin{cases} \dfrac{y^{r-1} e^{-y/2}}{2^r \Gamma(r)}, & H_0 \\ \dfrac{1}{2}\left(\dfrac{y}{2r\gamma}\right)^{r-1} e^{-\frac{2r\gamma + y}{2}} I_{r-1}(\sqrt{2r\gamma y}), & H_1 \end{cases} \quad (4)$$

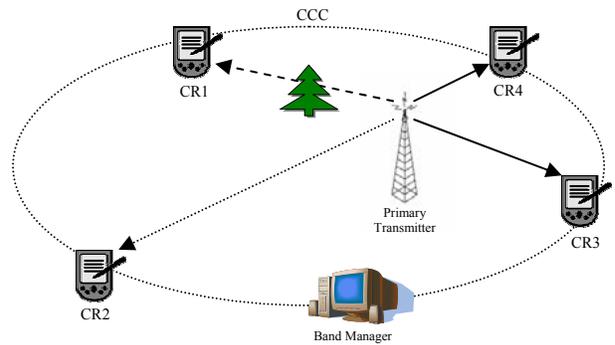

Fig. 2. Spectrum sensing scenario

where $\Gamma(.)$ is the gamma function and $I_v(.)$ is the modified Bessel function of the first kind of order $v$ [13].

Performance of energy detector is generally represented by the complementary receiver operating characteristics (cROC) curves (the plot of $P_m$ vs. $P_f$), where $P_m = 1 - P_d$ is defined as the probability of missed detection. In this context, the probability of detection ($P_d$) defines the sensing accuracy of primary transmitter hence measures the degree of interference protection to the primary receivers. The probability of false alarm ($P_f$) is the percentage of white spaces which are incorrectly sensed as occupied and a lower $P_f$ indicates higher spectrum utilization.

The cROCs for AWGN, Rayleigh fading and lognormal shadowing obtained in [8] clearly demonstrate that the performance of cooperative sensing results in better spectrum utilization and lower interference probability under fading or shadowing than individual sensing. While this underpins the importance of cooperation in successful sensing, further improvement is needed for reliable deployment of cognitive radios. In the following, we first briefly present cooperative sensing followed by our proposed method.

## III. COOPERATIVE SPECTRUM SENSING IN FADING/SHADOWING ENVIRONMENT

Fig. 2 depicts a spectrum sensing scenario for cognitive radio (CR) networks. Here, four users (CRs) are trying to detect the primary transmitter. But it is obvious that some might face shadowing and can confuse with white space while others, even all of them, might be subject to fading. But, as evidenced below, the possibility of sensing the primary transmitter as well as spectrum utilization increases much higher through cooperation.

Each user communicates with the band manager (secondary base station) through a common control channel (CCC) with either its measured energy, $Y$, or the function of it. Based on the received measurements, the band manager takes the final decision on the status of the band which is then broadcasted to all users. This decision can be taken in two ways – one is based on the measured energy of the users, namely *soft combining* and the other considers only the final one bit decision from each user, namely *hard combining*. Although soft combining needs only a slightly higher communication overhead to quantize the measured energy using a sufficient number of bits, its performance is much better compared to hard com-bining [6], [8]. This motivates the current research (presented details in Section IV & V) to focus on soft combining for better sensing ability.

Let $n$ be the number of cooperating users. For convenience, it is assumed that all $n$ users experience independent and identically distributed (iid) fading or shadowing. In soft combining, each of the users provides their measured signal energy to the band manager. Under these circumstances, the simplest method of combining is equal gain combining (EGC) which is investigated in recent works like [8] in the context of cognitive radio. Under EGC, the band manager decides between $H_0$ and $H_1$ by comparing the sum of measured energies to a threshold. The decision statistic is thus

$$Y_t = \sum_{i=1}^{n} Y_i \qquad (5)$$

and the output SNR, $\gamma_t$, is the sum of the SNRs ($\gamma_i$) on all the branches. Now, linear sum of $n$ iid noncentral $\chi^2$ variates ($Y_i$), each with $2r$ degrees of freedom and noncentrality parameter of $2r\gamma_i$, results in another $\chi^2$ variate ($Y_t$) with $2nr$ degrees of freedom and noncentrality parameter of $2r\gamma_t$ [12].

In a radio frequency environment where no fading exists, AWGN is the source of noise and channel gain $g$ is deterministic. Under this situation, using the cumulative distribution function of $Y$, the probabilities of detection $\Psi_d$ and false alarm $\Psi_f$ under cooperation can be expressed as

$$\Psi_d\big|_{AWGN} = P\{Y_t > \lambda \mid H_1\} = Q_{nr}\left(\sqrt{2r\gamma_t}, \sqrt{\lambda}\right) \qquad (6)$$

$$\Psi_f = P\{Y_t > \lambda \mid H_0\} = \frac{\Gamma(nr, \lambda/2)}{\Gamma(nr)} \qquad (7)$$

where $Q_{nr}(.,.)$ is the generalized Marcum $Q$ function [14], $\Gamma(.)$ and $\Gamma(.,.)$ are the complete and incomplete gamma functions respectively [13]. It is worth noting that as $\Psi_f$ does not depend on SNR, so expression of this parameter remains same throughout this paper, irrespective of the fading/shadowing condition and combining technique.

In the case of iid shadowing or fading, the probability of detection is conditioned on the instantaneous SNR, $\gamma$. As a consequence, the probability of detection may be derived by averaging (6) over fading statistics

$$\Psi_{de}\big|_{fading} = \int_{\gamma} Q_{nr}\left(\sqrt{2rx}, \sqrt{\lambda}\right) f_{\gamma_t}(x) dx \qquad (8)$$

where the subscript '$e$' in notation refer to EGC. The fading statistic can be different for different types of fading. Under *Rayleigh* fading, the pdf of $\gamma_t$ follows a gamma distribution [15] while under *Nakagami* fading, the fading statistic [16] includes $m$, the Nakagami parameter. Increasing value of $m$ decreases the amount of fading. Under *shadowing*, when $\gamma_t$ is log-normally distributed, the average probability of detection should be obtained numerically. In [8], analytical results for soft combining were presented under Rayleigh fading and lognormal shadowing. The authors, however, did not analyze probabilities under Nakagami fading.

One limitation of the EGC method, as investigated in [8], is that it gives equal weight to all the measured values which is not very effective in cognitive radio network where users may experience different degrees of fading or shadowing due to large separation in geographical area at any instant. Moreover, the user who is receiving higher power would be more reliable and thus be given higher weight. Considering these criteria, a new combining method for diverse primary transmitter signal energy detection in cognitive radio networks is proposed in the next section which overcomes the limitation of EGC..

IV. PROPOSED WEIGHTED COMBINING

In the following, we illustrate how different users may experience varied degree of fading or shadowing, or their locations w.r.t. primary transmitter may make one's measurement more reliable than others. For example, in Fig. 2 CR1 is subject to shadowing, so it receives lower signal power, CR2 receives lower signal power too, due to distance dependent path loss. So, in reaching a combined decision through cooperation, weight to each user's contribution should vary and depend on the received signal power as well as path loss. A method named weighted gain combining has been investigated in [11], but the effect of spatial variation of the users was not considered in it. Moreover, it considers average received power over some previous observations to give the weights, which may not be very effective in cognitive radio networks, where fast change in RF environment is very obvious.

Distance dependent path loss and received signal power are considered in our proposed method, called weighted combining (WC). Any user facing a higher path loss (e.g. CR2) than other (e.g. CR4) is given lower weight because its measurement may not be reliable enough to differentiate between the primary transmitter signal and locally generated unexpected noise. On the other hand, a user having higher received signal energy (e.g. CR3) than the other (e.g. CR2) should be given higher weight as it has better understanding of the primary transmitter. As the cognitive radio devices are considered to be position aware, they can easily get their distances from the primary transmitter. Let us consider, at a given instant, a user reports its current distance $d_i$ from the primary transmitter and the received signal energy $Y_i$. Then the band manager can obtain the weight factor $a_i$ for user $i$ by the following method

$$a_i = \frac{\gamma_i' \text{dB}}{L_i \text{dB}} = \frac{10\log(Y_i/Y_m)}{10\nu\log(d_i/d_m)} = \frac{10\log(\gamma_i/\gamma_m)}{10\nu\log(d_i/d_m)} \qquad (9)$$

where, $\nu$ is the path loss factor, $Y_m = \sum Y_i / n$ and $d_m$ is the mean distance of users from the primary transmitter. Following the argument above, the $a_i$ is taken as proportional to the dB normalized received energy and inversely proportional to the dB path loss w.r.t. primary transmitter (note that in this analysis primary is considered stationary). Considering the fact that $B$ and $T$ are same for each user, it can easily be deduced that $Y_i / Y_m = \gamma_i / \gamma_m$ by using (2). Thus, the combined output in WC scheme is

$$Y_t = \sum_{i=1}^{n} a_i Y_i \qquad (10)$$

We can express the decision statistic as

$$Y_t \sim \begin{cases} \chi^2_{2nr}, & H_0 \\ \chi^2_{2nr}(2r\gamma_t), & \text{where}, \gamma_t = \sum_{i=1}^{n} a_i \gamma_i \quad H_1 \end{cases} \quad (11)$$

This new expression of $\gamma_t$ is used into (6), (7) and (8) to evaluate the expression for detection probability ($\Psi_{dw}$) in WC, under different fading and shadowing while probability of false alarm $\Psi_f$ remains same as (7). In the following, we formulate probability of detection under different conditions in our proposed method.

*A. Rayleigh Fading*

In the case of iid Rayleigh fading, the pdf of $\gamma_t$ can be written as

$$f_{\gamma_t}(x) = \frac{x^{n-1} e^{-(x/\bar{\gamma}_w)}}{\Gamma(n) \bar{\gamma}_w^n} \quad (12)$$

where $\bar{\gamma}_w$ is the mean SNR for iid fading with WC and can be expressed by means of the virtual branch technique used in [18] as

$$\bar{\gamma}_w = \frac{\bar{\gamma}}{n} \sum_{i=1}^{n} a_i = \frac{a\bar{\gamma}}{n} \quad (13)$$

Now, by substituting (12) into (8), we have

$$\Psi_{dw}\big|_{ray} = \frac{1}{\Gamma(n) \bar{\gamma}_w^n} \int_{\gamma} Q_{nr}\left(\sqrt{2rx}, \sqrt{\lambda}\right) x^{n-1} e^{-(x/\bar{\gamma}_w)} dx \quad (14)$$

Recursive evaluation of this integral as described in [12] and use of the expressions from 0 yields

$$\Psi_{dw}\big|_{ray} = C_R + \left(\frac{n}{n+a r\bar{\gamma}}\right)^n e^{-\frac{\lambda}{2}} \quad (15)$$

$$\times \sum_{i=1}^{nr-1} \left(\frac{\lambda}{2}\right)^i \frac{1}{i!} \, {}_1F_1\left(n; i+1; \frac{\lambda}{2} \frac{a r\bar{\gamma}}{n+a r\bar{\gamma}}\right)$$

where ${}_1F_1(.;\,.\,;\,.)$ is the confluent hypergeometric limit function [13] and

$$C_R = \left(\frac{a r\bar{\gamma}}{n+a r\bar{\gamma}}\right) e^{-\frac{\lambda}{2} \frac{n}{n+a r\bar{\gamma}}} \left[\sum_{k=0}^{n-1} \zeta \left(\frac{n}{n+a r\bar{\gamma}}\right)^k \right.$$

$$\left. \times L_k\left(-\frac{\lambda}{2} \frac{a r\bar{\gamma}}{n+a r\bar{\gamma}}\right)\right] \quad (16)$$

where $L_v(.)$ is the Laguerre polynomial of degree $v$ [13]. The value of $\zeta = 1 + n/a r\bar{\gamma}$ when $k = n-1$, otherwise $\zeta = 1$.

*B. Nakagami Fading*

Nakagami fading is more generalized than Rayleigh fading and in this case (for iid) the pdf can be expressed as,

$$f_{\gamma_t}(x) = \left(\frac{m}{\bar{\gamma}_w}\right)^{mn} \frac{x^{mn-1}}{\Gamma(mn)} e^{-\left(\frac{m}{\bar{\gamma}_w}\right)x} \quad (17)$$

where $m$ denotes the Nakagami parameter. A similar method as in Rayleigh fading is used to calculate the detection probability which gives

$$\Psi_{dw}\big|_{nak} = \frac{m^{mn}}{\Gamma(mn) \bar{\gamma}_w^{mn}} \int_{\gamma} Q_{nr}\left(\sqrt{2rx}, \sqrt{\lambda}\right) x^{mn-1} e^{-\left(\frac{m}{\bar{\gamma}_w}\right)x} dx \quad (18)$$

Evaluation of this integral yields to the expression of the probability of detection under iid Nakagami fading. Ultimately, if all $n$ are replaced with $mn$ in (15) and (16), the closed form expression of $\Psi_{dw}\big|_{nak}$ can be obtained.

*C. Lognormal Shadowing*

Under lognormal shadowing, the pdf of SNR is

$$f_{\gamma_t}(x) = \frac{10/\ln 10}{\sqrt{2\pi\sigma^2}\gamma_t} e^{-\frac{(10\log\gamma - \mu)^2}{2\sigma^2}} \quad (19)$$

here $\mu = \overline{10\log\gamma}$ (dB) and $\sigma$ is the logarithmic standard deviation of shadowing in dB. As a result, if this distribution (19) is substituted into (8), $\Psi_{dw}$ has no closed form expression. The average probability of detection has to be computed numerically and thus the integral should be evaluated by using Gauss Hermite quadrature integration [19] to obtain the following expression

$$\Psi_{dw}\big|_{log} = \frac{1}{\pi} \sum_{i=1}^{l} w_i Q_{nr}\left[\sqrt{2am 10^{(x_i\sqrt{2\sigma}+\mu)/10}}, \sqrt{\lambda}\right] \quad (20)$$

The values $\{x_i\}$ and $\{w_i\}$, $i=1, 2, \ldots, l$ indicates the zeros and weights of the $l$-th order Hermite polynomial. The value of $l$ is chosen depending on the desired degree of accuracy. Its value is taken as 5 in evaluating performance to ensure sufficient accuracy of the result.

For $a_i=1$, for all $i$, the above expressions for WC reduces to those for EGC validating their correctness. Various performance analyses and comparison of WC and EGC under the above mentioned fading and shadowing environment is presented in the next section.

V. PERFORMANCE ANALYSIS

The performance of cooperative spectrum sensing is evaluated through its complementary ROC ($\Psi_m = 1-\Psi_d$ vs $\Psi_f$) curve, spectrum utilization and SNR requirements for different situations of interest. Fig. 3 illustrates the complementary ROC curves for WC and EGC under iid Rayleigh fading for different number of cooperating users, $n$. From this figure, it can generally be inferred that any type of cooperation improves the performance of spectrum sensing (as the curves shift towards the lower-left). It is observed that WC performs better than EGC irrespective of the number of cooperating users and its relative performance improves with increasing $n$.

For example, with $n=3$ and detection probability $\Psi_d = 0.9$ ($\Psi_m = 0.1$), probability of false alarm for EGC ($\Psi_{fe}$) is 0.063 while that for WC ($\Psi_{fw}$) is 0.02, which indicates a significant performance improvement by WC (68.25% over EGC). Similarly, for $\Psi_d = 0.99$ and $n=3$, the improvement is about 16% ($\Psi_{fe}=0.38$ and $\Psi_{fw}=0.32$). Relative improvement in both measures ($\Psi_d$ and $\Psi_f$) increases with increasing $n$ as evident by the increasing gap between the complementary ROC curves of both the schemes. The improvement is achieved due to the fact that WC scheme gives appropriate weight to user's contribution based on received power and path loss as opposed to equal weight in EGC.

The utilization of spectrum presented in Fig. 4 clearly depicts that RF spectrum is better utilized in WC as, for a given probability of detection, it achieves lower false alarm probability with the same number of users which, in turns, means that WC requires lower number of users than EGC to maintain the same spectrum utilization level. The mean SNR requirement to maintain detection and false alarm rate at a certain level are shown in Fig. 5 which confirms that WC requires less mean SNR with same number of cooperating users. Lower mean SNR requirement indicates better robustness of WC in higher noise uncertainty.

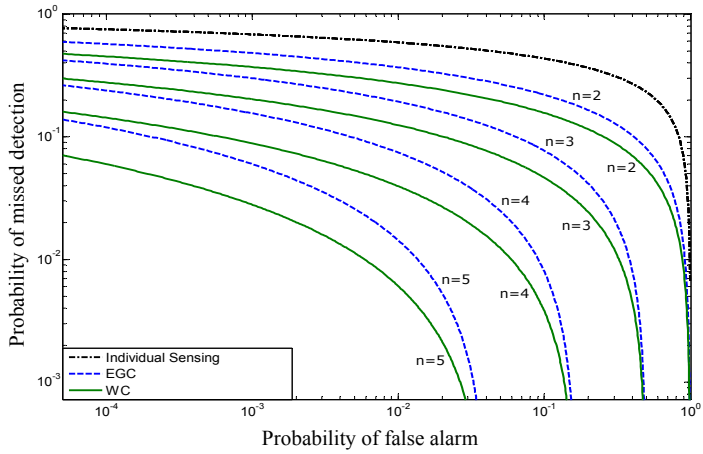

Fig. 3. Complementary ROC curve for iid Rayleigh fading for cooperative sensing varying *n*. $r=TW=1$, $\bar{\gamma}=6$ dB and $v=4$.

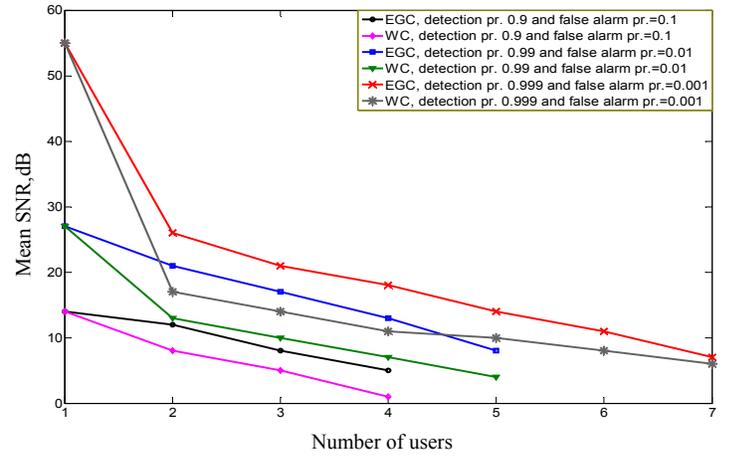

Fig. 5. SNR requirements curve for iid Rayleigh fading for cooperative sensing with various detection levels. $r=TW=1$ and $v=4$.

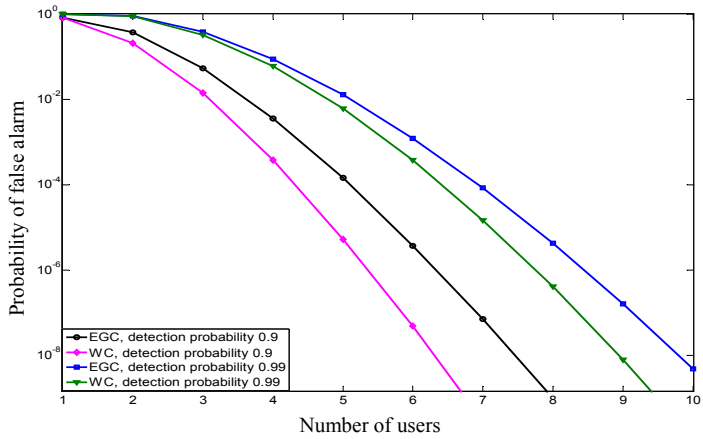

Fig. 4. Spectrum utilization curve for iid Rayleigh fading for cooperative sensing. $r=TW=1$, $\bar{\gamma}=6$ dB and $v=4$.

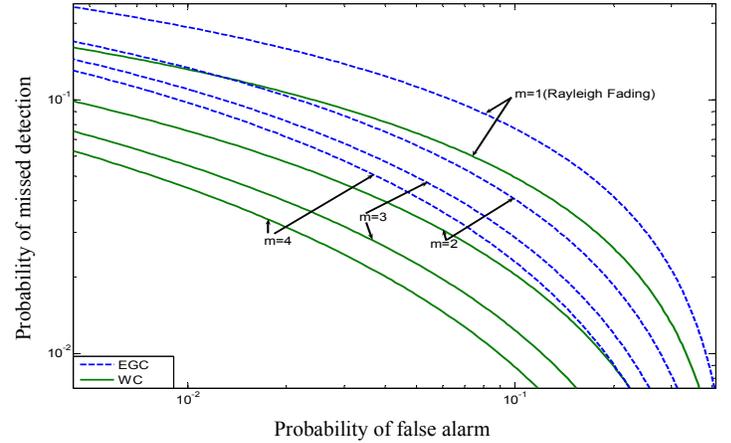

Fig. 6. Complementary ROC curve for iid Nakagami fading for cooperative sensing varying *m*. $r=TW=1$, $\bar{\gamma}=6$ dB, $v=4$ and n=3.

The complementary ROC curves in Fig. 6 indicate the performance superiority of WC over EGC in case of Nakagami fading. It is evident from the curve that, with same number of cooperating users, WC outperforms EGC in sensing accuracy for any degree of fading (*m*). For example, when $m=2$ and $\Psi_d=0.99$, the improvement is about 35.71%, similar improvement is observed for m=3 &4. Results (not presented here) indicate that WC also outperforms EGC in spectrum utilization and SNR requirements for Nakagami fading.

Performance analyses of WC for lognormal shadowing are illustrated in Fig. 7~8. Fig. 7 shows that, at $\Psi_d = 0.9$ and $n=3$, $\Psi_{fw}=0.0667$ while $\Psi_{fe}=0.1273$ which indicates an improvement of 47.60% by WC. But the improvement rises to 68.98%

($\Psi_{fe}=0.0033$ and $\Psi_{fw}=0.001$) when $n=5$. The spectrum utilization curves in Fig. 8 illustrates that WC utilizes the spectrum more efficiently than EGC for lower probability of detection. However, when higher detection accuracy need to be achieved, relative performance of WC becomes trivial but yet it maintains its superiority over EGC.

## VI. Concluding Remarks

In this paper we proposed an improved cooperative-sensing-based opportunistic spectrum access under fading and shadowing, and theoretically formulated probability of detection and false alarm rate for each case. As indicated by the presented analyses, user collaboration in the proposed way result in significant enhancements in detection and spectrum utilization over previously practiced EGC method. By employing weight to the decision of individual sensing, WC can diminish the probability of missing white spaces while providing the primary receiver with its desired level of interference-protection using less number of cooperating users than existing scheme. Moreover, by considering the spatial variation of users, it represents the real scenario more accurately. Although WC requires higher number of bits to inform the band manager about the measured energy and the position, this overhead is alleviated by incorporating much less number of users. With its improved performance, the proposed sensing scheme will facilitate adoption of cognitive radio networks.

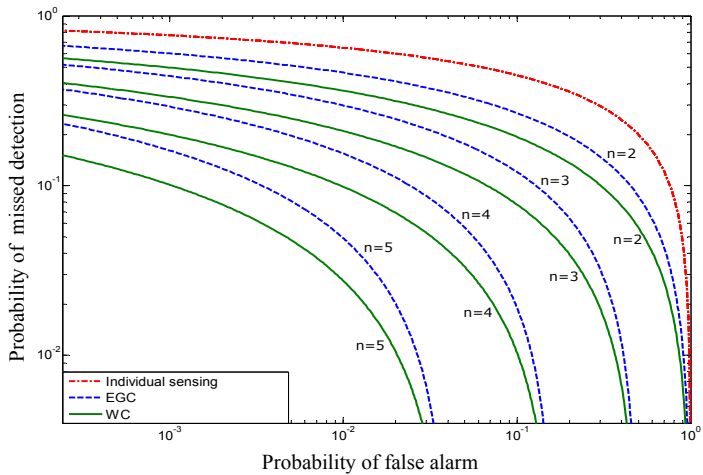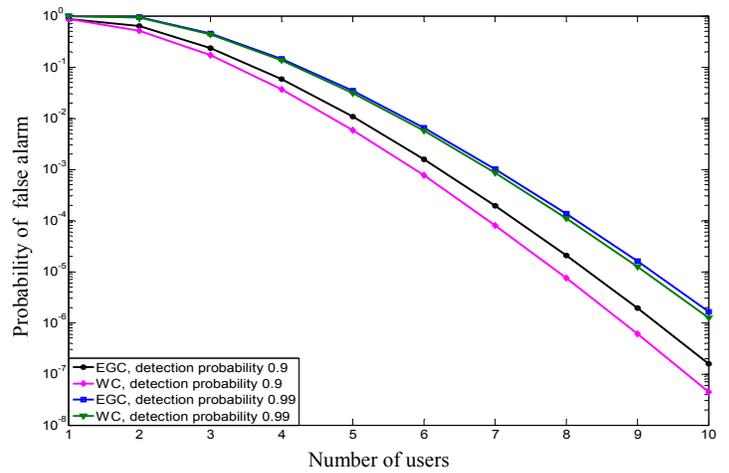

Fig. 7. Complementary ROC curve for iid lognormal shadowing for cooperative sensing varying $n$. $r=TW=1$, $\mu=1$ dB, $\sigma=6$dB and $v=4$.

Fig. 8. Spectrum Utilization curve for iid lognormal shadowing for cooperative sensing. $r=TW=1$, $\mu=1$ dB, $\sigma=6$dB and $v=4$.